# Sensing remote nuclear spins

Nan Zhao[1], Jan Honert[1], Berhard Schmid[1], Junichi Isoya[2], Mathew Markham[5], Daniel Twitchen[5], Fedor Jelezko[3], Ren-Bao Liu[4], Helmut Fedder[1], Jörg Wrachtrup[1]

1. 3rd Institute of Physics and Research Center SCOPE, University Stuttgart, Pfaffenwaldring 57, 70569 Stuttgart, Germany
2. Graduate School of Library, Information and Media Studies, University of Tsukuba, 1-2 Kasuga, Tsukuba, Ibaraki 305-8550, Japan
3. Institut für Quantenoptik, Universität Ulm, 89081 Ulm, Germany
4. Department of Physics, The Chinese University of Hong Kong, Shatin, New Territories, Hong Kong, China
5. Element Six Ltd, Ascot SL5 8BP, Berks, England

**Sensing single nuclear spins is a central challenge in magnetic resonance based imaging techniques. Although different methods and especially diamond defect based sensing and imaging techniques in principle have shown sufficient sensitivity, signals from single nuclear spins are usually too weak to be distinguished from background noise. Here, we present the detection and identification of remote single $^{13}$C nuclear spins embedded in nuclear spin baths surrounding a single electron spins of a nitrogen-vacancy centre in diamond. With dynamical decoupling control of the centre electron spin, the weak magnetic field ~10 nT from a single nuclear spin located ~3 nm from the centre with hyperfine coupling as weak as ~500 Hz is amplified and detected. The quantum nature of the coupling is confirmed and precise position and the vector components of the nuclear field are determined. Given the distance over which nuclear magnetic fields can be detected the technique marks a firm step towards imaging, detecting and controlling nuclear spin species external to the diamond sensor.**

By continuously improving sensitivity and spatial resolution of magnetic field imaging techniques, it became possible to detect single electron spins [1] or small ensemble of nuclear spins [2]. Further increasing sensitivity would allow for single nuclear spin detection with unprecedented impact on structure analysis ranging from life science to material research as a whole [3,4]. In addition, nuclear spins are valuable resources for quantum registers particularly in the case of diamond defects [5-7]. Being able to detect as remote as possible nuclear magnetic fields and isolate them from surrounding noise fields is a key requirement in both fields.

Detection of single nuclear spins remains challenging, due to their extremely weak signals in comparison with the background noise sources. At present, only the detection of nuclear spins which are strongly coupled to e.g. diamond defect sensor electrons is possible [5, 8]. This strong coupling requirement significantly limits the detection range of single nuclear spins, and reduces the possibility of imaging external nuclear spin as well as using them e.g. for quantum memories. The situation will be greatly improved if we could detect single nuclear spins far away and weakly coupled to diamond defects. The signal of weakly coupled nuclear spins usually is easily drowned in background fields leading to reduced sensor spin sensitivity revealed by a reduced sensor spin dephasing time $T_2^*$ [8]. In this work, we overcome this $T_2^*$ limitation, and demonstrate the detection of remote, weakly coupled single nuclear spins by employing coherent averaging techniques on the sensor electron spin while retaining sensitivity to specific nuclear magnetic fields.

As sensor spin we use the electron spins of a single nitrogen-vacancy (NV) centre in isotope purified diamond CVD sample ($^{12}C$ abundance > 99.99%, see Fig. 1a for the structure of NV centre and a schematic illustration of nuclear spin bath). In this sample, the typical distance of the nearest $^{13}C$ nuclear spin to the NV centre is ~ 3 nm. At such distances, single nuclear spins have weak coupling (~ 1 kHz) to the centre electron spin. Since the detected nuclear spin is embedded in a nuclear spin bath which produces background noise, the transition frequency change of the centre spin due such a small coupling cannot be observed straight forward e.g. by a free induction decay (FID) or Ramses fringe experiment. Figure 1b shows a typical FID of the NV centre spin coherence in the thermal noise of $^{13}C$ bath spins. The coherence decays within a dephasing time $T_2^*$ ~200 μs, which limits the sensitivity to nuclear spin couplings to be less than ~ 5 kHz. In order to overcome the $T_2^*$ limit, and to distinguish weak single nuclear spin signal from background noise, we need a tool with simultaneously high sensitivity to weak signals and ability to filter our background noise.

Dynamical decoupling control [9] of electron spins is an ideal tool which meets the both requirements [3]. In particular, the widely used Carr-Purcell-Meiboom-Gill (CPMG) control sequence [10-12] selectively amplifies the signal at specific frequency, and suppresses unwanted background noise (see Fig. 1c for the pulse sequence of CPMG). With these unique features of the CPMG control sequence, noise spectra of single quantum objects, like superconducting qubits [13], trapped ions [14], have been successfully obtained. In these works, the detected noise spectra are structureless, in sharp difference to the nuclear spin bath. The dynamics of single nuclear spins causes peak structures on the magnetic field noise spectrum surrounding the NV sensor spin [3]. We use CPMG control of the electron spin to resolve this fine-structure and to isolate single nuclear spin signals from the bath.

The decoherence behavior $L(t)$ of a qubit under dynamical decoupling control is usually determined by the noise spectrum $S(\omega)$ as [15]

$$L(t) = \exp[-\chi(t)] = \exp\left[-\int_0^t \frac{d\omega}{2\pi} \frac{S(\omega)}{\omega^2} F(\omega; t)\right], \qquad (1)$$

where $F(\omega;t)$ is the filter function associated with the control pulse scheme (see Fig. 1d for the filter function of CPMG sequence). In our sample, the centre electron spin coherence time is $T_2$~3.0 ms under Hahn echo control, which is mainly limited by the spin-lattice relaxation ($T_1$ process) and residual noise from the electron spin bath [10]. By increasing the CPMG control pulse number, electron spin coherence is well protected with a plateau in $L(t)$ for $t < 1$ ms [16].

Nuclear spins around the NV centre bring additional structures to the smooth noise spectrum (see Fig. 1d). The precession of nuclear spins produces a peak in the spectrum centered at the nuclear spin Larmor frequency $2\pi f_L = \gamma_{nuc} B$ with $\gamma_{nuc}$ being the nuclear spin gyromagnetic ratio. This peak structure exhibits itself as a coherence dip according to Eq. (1). Figures 2a-2c show the coherence dips of different NV centres in various magnetic field strength $B$ and under $N$-$\pi$-pulse control (CPMG-$N$). The strong peaks of the filter function $F(\omega,t)$ in Eq. (1) give rise to the coherence dips at

$$t_{dip} = \frac{(2k-1)N}{2 f_L}. \qquad (2)$$

The dependence of the dip position on the dip order $k$, control pulse number $N$ and magnetic field strength $B$ is shown in Fig. 2c, which allows to determine the gyromagnetic ratio $\gamma_{nuc} = \gamma_C = 2\pi \times 1.07$ kHz/G. Thus, we identify the origin of the coherence dips as the precession of $^{13}$C nuclear spins. Furthermore, the depths of coherence dips reveal the signal amplitudes form nuclear spins. Notice that the integral $\chi(t)$ in Eq. 1 is $\chi(t) \sim N^2 (A_j^\perp)^2 / B^2$ [3], where $A_j^\perp$ is the orthogonal component of the hyperfine coupling $\mathbf{A}_j$ of the $j$th nuclear spin. With $\chi(t) \sim 1$ in the present case, we estimate the hyperfine coupling amplitude $A_j^\perp \sim |\mathbf{A}_j| \sim B/N \sim 1$ kHz, which is consistent with the average $^{13}$C abundance of the sample.

Two qualitatively distinct mechanisms could cause similar coherence dips shown in Fig. 2, namely, *classical* mechanism and *quantum* mechanism [17-19]. On the one hand, a large number of *incoherent* $^{13}$C nuclear spins precessing with the same frequency but with random phases and amplitudes impose fluctuating magnetic field at the NV centre, inducing the coherence dips as described by Eq. 1. This process is of classical nature, since it is essentially equivalent to the case of artificially generated

**4**

classical AC-magnetic field [20-22]. On the other hand, *coherent* coupling to single nuclear spins can also produce electron spin coherence modulation [3, 5], which is a quantum mechanism since the quantum phase of nuclear spin is well-preserved in the process. In the following, we will rule out the classical mechanism and identify quantum coherent coupling of single nuclear spins as the leading mechanism.

Increasing the CPMG control pulse number yields additional information about the nuclear spin bath. Figure 3a shows the coherence dips of NV-B under CPMG controls of different pulse number $N=30$ and $N=60$. As expected by the classical theory (see Eq. 1 and the filter function of CPMG), the depth of the dip increases upon increasing the pulse number $N$. However, dips become negative for $N=60$ (Fig. 3a). This is in strong contradiction of Eq. 1, in which the coherence is always $\geq 0$. The negative dips in Figs. 3 and 4 however are well explained by the quantum mechanism. In the quantum decoherence picture, the coherent coupling between electron and nuclear spins creates entanglement, and electron spin coherence is measured by the overlap between the two different nuclear spin states as [23-26],

$$L(t) = \langle J_0(t) | J_\pm(t) \rangle, \qquad (3)$$

where $|J_{0,\pm}(t)\rangle$ is the nuclear spin state at time $t$ depending on centre electron spin state $|m_S = 0\rangle$ and $|m_S = \pm 1\rangle$. This quantum entanglement induced decoherence is bounded between -1 and 1 (Fig. 3a).

The quantum nature of the coherence dip is also proved by the quantum back-action effect [17-19]. Whether the noise dynamics is affected by the qubit state is an important criterion to distinguish the quantum or classical nature of the bath [17-19]. In additional to external controls (e.g., the applied magnetic fields), quantum evolution of bath spins is also influenced by the centre spin states (i.e., the back-action), and results in different decoherence behavior for the centre spin being prepared in different states. In contrast, decoherence due to the classical noise is not sensitive to the centre spin state. We use the coherence of the two electron spin transitions (i.e., $|m_S = 0\rangle \leftrightarrow |m_S = \pm 1\rangle$) to check the quantum back-action effect. Figures 3a & 3b show the coherence dips of NV-A and NV-B for both transitions. The coherence dips of the two transitions appear at different times with a clear shift between the two transitions. The magnitude of the shift linearly increases upon increasing on the pulse number $N$ or the dip order $k$, as summarized in Fig. 3d.

The relative shift of the coherence dips for $|m_S = 0\rangle \leftrightarrow |m_S = \pm 1\rangle$ transitions is explained by the modified Larmor frequency due to the quantum back-action. To be specific, the back-action to $^{13}$C nuclear spin $\mathbf{I}_j$ is givn by the effective hyperfine field $\mathbf{A}_j$. For the NV centre electron spin prepared in the superposition of $|m_S = 0\rangle$ and $|m_S = +1\rangle$ states, the intrinsic Larmor frequency of spin $\mathbf{I}_j$ is shifted due to the hyperfine field component $A_j^z$ (the component parallel to the external field $\mathbf{B}$). While, for the NV centre electron spin in $|m_S = 0\rangle$ and $|m_S = -1\rangle$ superposition states, the Larmor frequency is modified by the same amplitude but with an opposite sign. Thus different effective Larmor frequencies cause the different coherence dip positions resolved in the CPMG measurements.

With this observation, we can obtain the parallel hyperfine field component $A_j^z$ from the dip position shifts. In our experiments, the relative change of Larmor frequency due to quantum back-action is small (i.e., $|A_j^z| \ll B$). In this case, the dip position shift is $\delta t_{dip} = (|A_j^z|/B)t_{dip}$ ($t_{dip}$ being the time determined solely by applied magnetic field, see Eq. 2). The parallel component $A_j^z$ is extracted from $\delta t_{dip}$ as shown in Fig. 3d. Furthermore, the orthogonal component $A_j^\perp$ determines the depth of the dip, whose value can be obtained by fitting the data with Eq. 3. Using this method, the hyperfine coupling strength is determined. The sensitivity is limited by the centre electron spin coherence time $T_2$. Making use of the long $T_2$ time ~3 ms, we have successfully observed the coherence dip at $t_{dip}$=2.4 ms with CPMG-60 (Fig. 4g). This dip is caused by a single $^{13}$C nuclear spin with coupling strength of about 500 Hz, corresponding to the distance of about 3.5 nm from the NV centre. This detection range is about 10 times larger than the $T_2^*$ limited scheme, and the sensitivity is improved by 3 orders of magnitudes [5].

With the understanding of the quantum nature of coherence dips, we can resolve more clearly the microscopic structure of the nuclear spin bath surrounding the electron spin. Figure 4 shows the coherence dip structures of NV-A. With increasing the pulse number, instead of a single sharp dip as shown in Fig. 3a, we observe dip splitting. Such splitting is caused by the simultaneous coherent coupling to two $^{13}$C nuclear spins in the spin bath. With a two-nuclear-spin model, the observed data is perfectly reproduced by the theory. The hyperfine coupling strengths for the two $^{13}$C spins are obtained from the model as $|A_1| = 2.9 \text{ kHz}$ and $|A_2| = 3.3 \text{ kHz}$, respectively. Thus, a difference in hyperfine coupling as small as 400 Hz can be resolved. By changing the direction of the applied magnetic field **B**, we can obtain hyperfine field projections along different directions. In this way, the *complete* vector components of the hyperfine coupling of the single nuclear spins and the position of the nuclear spin can be derived from experiments.

In summary, we have demonstrated an ultra-sensitive detection of the quantum fluctuation from remote single nuclear spins. We overcome the $T_2^*$ limit of conventional detection scheme, and achieve the detection of single nuclear spins with coupling strength as weak as 500 Hz relating to a distance of ~3.5 nm and are able to resolve two nuclear spins differing in hyperfine coupling by only 400 Hz. The significant enhancement of sensitivity and sensitivity will enable the detection of external single nuclear spins outside, or on the surface of diamond sample by using NV centres close to the diamond surface [27]. The identification of single nuclear spins at a distance of nanometer thus opens the door to the single nuclear spin magnetic resonance and imaging [3]. We point out that our detection scheme of single nuclear spins is a fully quantum mechanical effect, which thus can be combined with sophisticated nuclear spin control techniques well-known from NMR. Our work also extends the number of quantum spin qubits around a diamond electron spin qubit, and hence may promote further investigation of using single weakly coupled nuclear spins as quantum registers in quantum information processing.

Authors' Note: During preparation of this manuscript, we got aware of a similar investigation [28].

# References

1. Rugar, D., Budakian, R., Mamin, H. J., & Chui, B. W. Single spin detection by magnetic resonance force microscopy. *Nature* **430**, 329 (2004).
2. Degen, C. L., Poggio, M., Mamin, H. J., Rettner, C. T., & Rugar, D. Nanoscale magnetic resonance imaging. *Proceedings of the National Academy of Sciences of the United States of America* **106**, 1313 (2009).
3. Zhao, N., Hu, J. L., Ho, S. W., Wan, J. T. K., & Liu, R. B. Atomic-scale magnetometry of distant nuclear spin clusters via nitrogen-vacancy spin in diamond. *Nature nanotechnology* **6**, 242 (2011).
4. Cai, J.-M., Jelezko, F., Plenio, M. B., Retzker. A., arXiv:1112.5502v1 (2011).
5. Childress, L., Gurudev Dutt, M. V., Taylor, J. M., Zibrov, A. S., Jelezko, F., Wrachtrup, J., Hemmer, P. R., et al. Coherent dynamics of coupled electron and nuclear spin qubits in diamond. *Science* **314**, 281 (2006)
6. Dutt, M. V. G., Childress, L., Jiang, L., Togan, E., Maze, J., Jelezko, F., Zibrov, A. S., et al. Quantum register based on individual electronic and nuclear spin qubits in diamond. *Science* **316**, 1312 (2007).
7. Neumann, P., Mizuochi, N., Rempp, F., Hemmer, P., Watanabe, H., Yamasaki, S., Jacques, et al. Multipartite entanglement among single spins in diamond. *Science* **320**, 1326 (2008)
8. Dréau, a., Maze, J.-R., Lesik, M., Roch, J.-F., & Jacques, V. High-resolution spectroscopy of single NV defects coupled with nearby $^{13}$C nuclear spins in diamond. *Physical Review B* **85**, 134107 (2012).
9. Viola, L., Knill, E., & Lloyd, S. Dynamical Decoupling of Open Quantum Systems. *Physical Review Letters* **82**, 2417, (1999).
10. de Lange, G., Wang, Z. H., Riste, D., Dobrovitski, V. V., & Hanson, R. Universal Dynamical Decoupling of a Single Solid-State Spin from a Spin Bath. *Science* **330**, 60, (2010).
11. Ryan, C., Hodges, J., & Cory, D. Robust Decoupling Techniques to Extend Quantum Coherence in Diamond. *Physical Review Letters* **105**, 200402 (2010).
12. Naydenov, B., Dolde, F., Hall, L. T., Shin, C., Fedder, H., Hollenberg, L. C. L., Jelezko, F., et al. Dynamical decoupling of a single-electron spin at room temperature. *Physical Review B*, **83**, 081201 (2011).
13. Bylander, J., Gustavsson, S., Yan, F., Yoshihara, F., Harrabi, K., Fitch, G., Cory, D. G., et al. Noise spectroscopy through dynamical decoupling with a superconducting flux qubit. *Nature Physics* **7**, 565 (2011).
14. Almog, I., Sagi, Y., Gordon, G., Bensky, G., Kurizki, G., DavidsonJ. N., *Phys. B: At. Mol. Opt. Phys.* **44**, 154006 (2011).
15. Cywiński, Ł., Lutchyn, R., Nave, C., & Das Sarma, S. How to enhance dephasing time in superconducting qubits. *Physical Review B* **77**, 174509 (2008).
16. Pham, L. M., Bar-Gill, N., Belthangady, C., Le Sage, D., Cappellaro, P., Lukin, M. D., Yacoby, A., & Walsworth, R. L. Enhanced solid-state multi-spin metrology using dynamical decoupling. arXiv:1201.5686v2 (2012).
17. Zhao, N., Wang, Z.-Y., & Liu, R.-B. Anomalous Decoherence Effect in a Quantum Bath. *Physical Review Letters* **106**, 217205 (2011).
18. Huang, P., Kong, X., Zhao, N., Shi, F., Wang, P., Rong, X., Liu, R.-B., et al. Observation of

an anomalous decoherence effect in a quantum bath at room temperature. *Nature Communications* **2**, 570 (2011).
19. Reinhard, F., Shi, F.-Z., Zhao, N., Rempp, F., Naydenov, B., Meijer, J., Hall, L. T., Hollenberg, L., Du, J.-F., Liu R.-B. & Wrachtrup, J. *Physical Review Letters*, to be published (2012).
20. Maze, J. R., Stanwix, P. L., Hodges, J. S., Hong, S., Taylor, J. M., Cappellaro, P., Jiang, L., et al. Nanoscale magnetic sensing with an individual electronic spin in diamond. *Nature* **455**, 644 (2008).
21. Balasubramanian, G., Chan, I. Y., Kolesov, R., Al-Hmoud, M., Tisler, J., Shin, C., Kim, C., et al. Nanoscale imaging magnetometry with diamond spins under ambient conditions. *Nature* **455**, 648 (2008).
22. de Lange, G., Ristè, D., Dobrovitski, V. V., & Hanson, R. Single-Spin Magnetometry with Multipulse Sensing Sequences. *Physical Review Letters* **106**, 080802 (2011).
23. Yao, W., Liu, R.-B., & Sham, L. Restoring Coherence Lost to a Slow Interacting Mesoscopic Spin Bath. *Physical Review Letters* **98**, 077602 (2007)
24. Witzel, W. M. & Das Sarma, S. Multiple-Pulse Coherence Enhancement of Solid State Spin Qubits, *Phys. Rev. Lett.* **98**, 077601 (2007).
25. Maze, J. R., Taylor, J. M., & Lukin, M. D. Electron spin decoherence of single nitrogen-vacancy defects in diamond. *Physical Review B* **78**, 94303 (2008).
26. Zhao, N., Ho, S.-wah, & Liu, R.-bao. Decoherence and dynamical decoupling control of nitrogen vacancy center electron spins in nuclear spin baths. *Physical Review B* 85, 115303 (2012).
27. Ofori-Okai, B. K., Pezzagna, S., Chang, K., Schirhagl, R., Tao, Y., Moores, B. A., Groot-Berning, K., Meijer, J., & Degen, C. L., Spin Properties of Very Shallow Nitrogen Vacancy Defects in Diamond. arXiv:1201.0871v1 (2012).
28. Kolkowitz, S., Unterreithmeier, Q.P., Bennett, S.D., Lukin, M.D., Sensing distant nuclear spins with a single electron spin. arXiv:1204.5483 (2012)

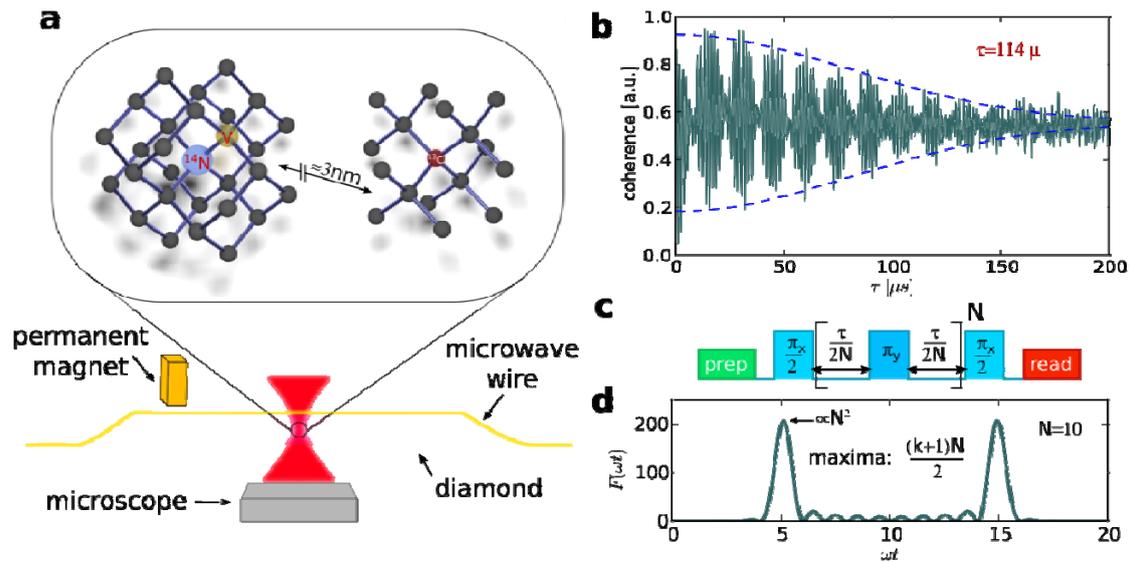

**Figure 1 | NV centre spin and $^{13}$C nuclear spin bath in diamond. a**, Schematic illustration of the confocal microscopy setup and NV centre in $^{13}$C purified diamond sample. The typical distance from NV centre to the nearest $^{13}$C is ~ 3 nm. **b**, A typical measurement of free-induction decay of NV centre coherence double quantum transition (i.e., between $|m_S = +1\rangle$ and $|m_S = -1\rangle$ states) with $T^*_{2,DQT}$=114 µs, corresponding to dephasing time $T^*_{2,SQT}$=228 µs for single quantum transition between $|m_S = 0\rangle$ and $|m_S = +1\rangle$ states). **c**, CPMG pulse sequence for single nuclear spin detection. **d**, The filter function of *N*-pulse CPMG control sequence. Strong peaks appear at $\omega t = (2k-1)N$.

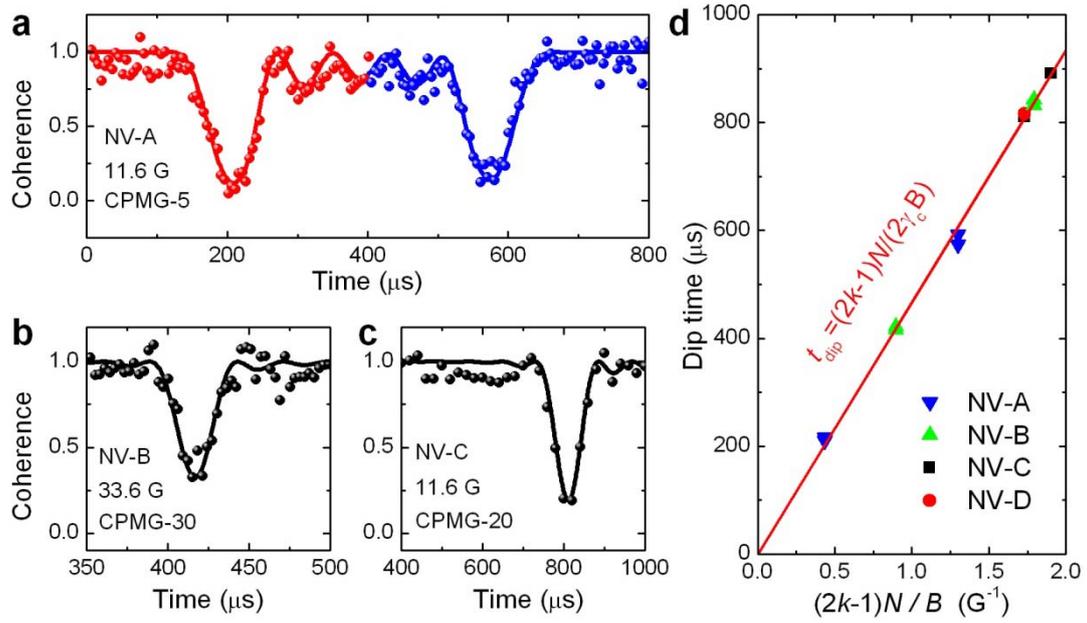

**Figure 2 | $^{13}$C nuclear spin induced coherence dips. a**, coherence dips of NV-A under magnetic field $B$=11.6 G and CPMG-5 control. The red and blue symbols corresponds to the 1$^{st}$ and 2$^{nd}$ order dips ($k$=1 and 2), respectively. **b**, the 1$^{st}$ order ($k$=1) coherence dip of NV-B under magnetic field $B$=33.6 G and CPMG-30 control. **c**, the 1$^{st}$ order ($k$=1) coherence dip of NV-C under magnetic field $B$=11.6 G and CPMG-20 control. Solid curves in **a**, **b**, and **c** are theoretical calculations according to Eq. 3. **d**, Coherence dip time as function of dip order $k$, control pulse number $N$, and magnetic field strength $B$ for various NV centres (different symbols). The red curve is the calculated dip time using $^{13}$C gyromagnetic ratio $\gamma_C=2\pi\times 1.07$ kHz/G.

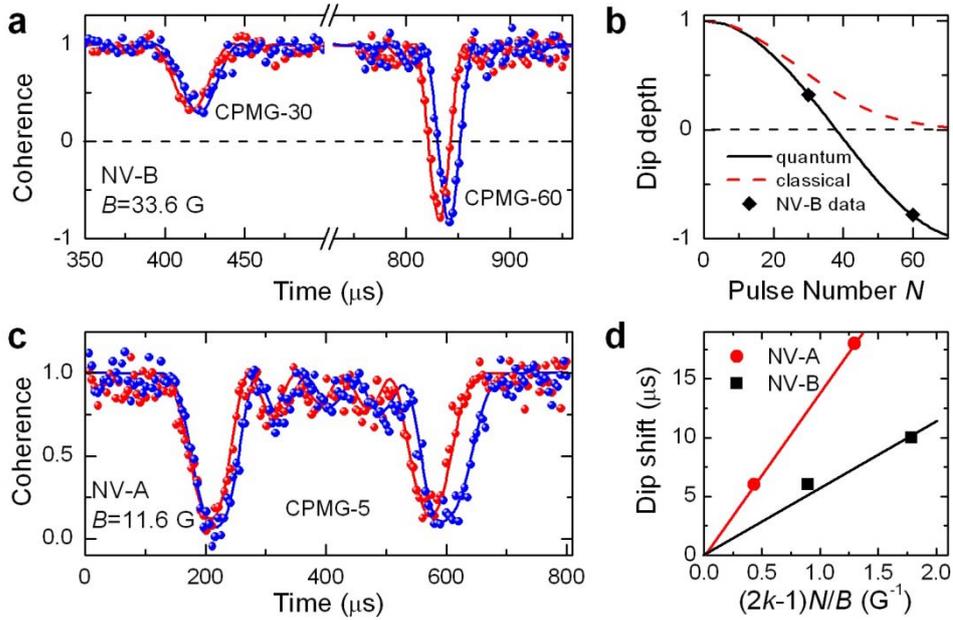

**Figure 3 | Quantum decoherence effect due to $^{13}$C nuclear spins. a**, Coherence dips of NV-B under CPMG-30 (left shallow dips) and CPMG-60 (right deep dips) control. Red and blue symbols are the measured coherence for $|0\rangle \leftrightarrow |-1\rangle$ and $|0\rangle \leftrightarrow |+1\rangle$ transitions, respectively. Solid curves are the corresponding theoretical calculations according to Eq. 3. **b**, Calculated dip depth of NV-B according to classical (dashed curve) and quantum mechanical (solid curve) decoherence mechanisms, respectively. Symbols are measured data from **a**. **c**, Identical to **a**, but for the 1$^{st}$ and 2$^{nd}$ coherence dips of NV-A under CPMG-5 control. **d**, Linear dependence of the relative dip shifts on the dip order $k$, control pulse number $N$, and magnetic field strength $B$. Symbols are measured data of NV-A and NV-B. Straight lines are fits with slopes determined by the parallel components of the hyperfine coupling.

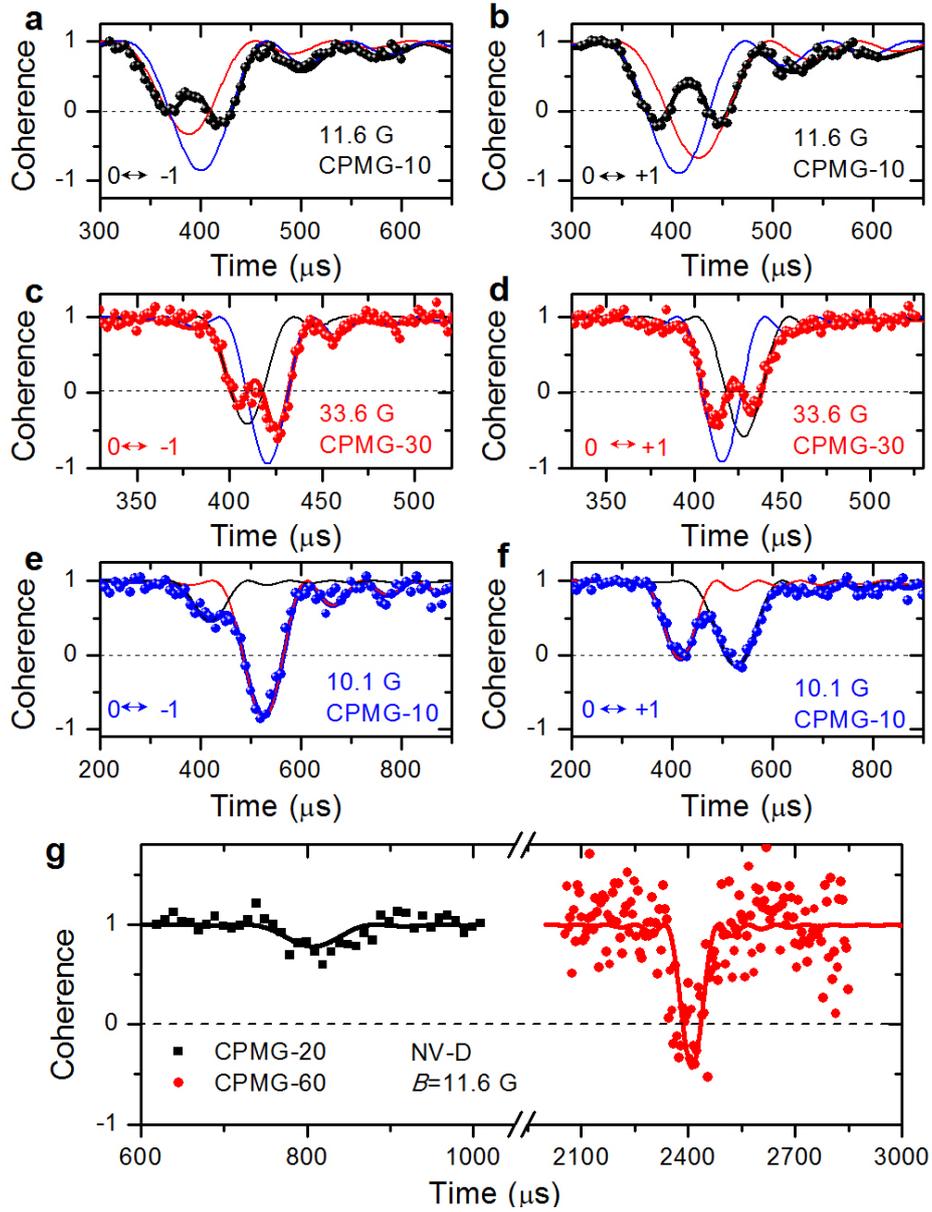

**Figure 4 | Determination of hyperfine coupling of single nuclear spins. a**, Coherence of $|0\rangle \leftrightarrow |+1\rangle$ transition of NV-A under CPMG-10 control and magnetic field $B$=11.6 G with the polar angle $\theta$=12.9° and the azimuth angle $\varphi$=135.0°. Symbols are measured data, the thin curves are calculated individual contribution of two $^{13}$C nuclear spins, and the thick black curve is the total contribution of the two spins. **b**, Identical to **a**, but for $|0\rangle \leftrightarrow |-1\rangle$ transition. The hyperfine coupling components extracted from **a** & **b** are $(A_1^z, A_1^\perp) = (1.2, 2.6)$kHz and $(A_2^z, A_2^\perp) = (0.2, 3.3)$kHz. **c** & **d**, The same as **a** &

**b**, but for magnetic field $B$=33.6 G with the polar angle $\theta$=29.4° and the azimuth angle $\varphi$=50.5°. The hyperfine coupling components are $(A_1^z, A_1^\perp) = (1.6, 2.5)$ kHz and $(A_2^z, A_2^\perp) = (-0.4, 3.3)$ kHz. **e** & **f**, The same as **a** & **b**, but for magnetic field $B$=10.1 G with the polar angle $\theta$=64.1° and the azimuth angle $\varphi$=81.5°. The hyperfine coupling components are $(A_1^z, A_1^\perp) = (2.5, 1.45)$ kHz and $(A_2^z, A_2^\perp) = (-2.6, 2.2)$ kHz. The hyperfine coupling magnitudes of the two $^{13}$C nuclear spins are consistently determined from **a-f** as $|A_1| = 2.9$ kHz and $|A_2| = 3.3$ kHz. **g**, CPMG-20 and CPMG-60 measurements on NV-D in the same magnetic field as that in **a** & **b**. Symbols are measured data. Solid curves are theoretical calculated dips with $(A^z, A^\perp) = (0.4, 0.15)$ kHz.